# Epitaxial Growth of Two-Dimensional Stanene

Fengfeng Zhu[1†], Wei-jiong Chen[1†], Yong Xu[2,3†], Chun-lei Gao[1,4], Dan-dan Guan[1,4], Canhua Liu[1,4], Dong Qian[1,4*], Shou-Cheng Zhang[2,3], Jin-feng Jia[1,4*]

[1]Key Laboratory of Artificial Structures and Quantum Control (Ministry of Education), Department of Physics and Astronomy, Shanghai Jiao Tong University, Shanghai 200240, China
[2]Department of Physics, McCullough Building, Stanford University, Stanford, California 94305-4045, USA
[3]Department of Physics and Institute for Advanced Study, Tsinghua University, Beijing 100084, China
[4]Collaborative Innovation Center of Advanced Microstructures, Nanjing 210093, China
[†]contributed equally to this work.
[*]Corresponding: dqian@sjtu.edu.cn; jfjia@sjtu.edu.cn

**Ultrathin semiconductors present various novel electronic properties. The first experimental realized two-dimensional (2D) material is graphene. Searching 2D materials with heavy elements bring the attention to Si, Ge and Sn. 2D buckled Si-based silicene was realized by molecular beam epitaxy (MBE) growth[1,2]. Ge-based germanene was realized by mechanical exfoliation[3]. Sn-based stanene has its unique properties. Stanene and its derivatives can be 2D topological insulators (TI) with a very large band gap as proposed by first-principles calculations[4], or can support enhanced thermoelectric performance[5], topological superconductivity[6] and the near-room-temperature quantum anomalous Hall (QAH) effect[7]. For the first time, in this work, we report a successful fabrication of 2D stanene by MBE. The atomic and electronic structures were determined by scanning tunneling microscopy (STM) and angle-resolved photoemission spectroscopy (ARPES) in combination with first-principles calculations. This**

**work will stimulate the experimental study and exploring the future application of stanene.**

The researches of two-dimensional (2D) materials, as inspired by the great success of graphene, have experienced an explosive increase in the recent years[8]. Particularly the 2D group IV materials that include graphene, silicene, germanene and stanene as a family have attracted enormous interests due to their exotic electronic properties. In this 2D material family, stanene is of special interest due to its outstanding properties. For instance, stanene and its derivatives could support large-gap 2D quantum spin Hall (QSH) states and thus enable dissipationless electric conduction at room temperature[4,9]. Moreover, stanene could also provide other novel features, including enhanced thermoelectricity[5], topological superconductivity[6] and near-room-temperature quantum anomalous Hall (QAH) effect[7]. Any of these theoretical proposals of stanene, if confirmed experimentally, could offer great opportunities not only for the development of condensed matter physics and material science but also for future technologies.

However, the experimental growth of stanene remains elusive. Various techniques have been developed for fabricating 2D materials, like mechanical exfoliation, CVD, and MBE. And thus most of the 2D group IV materials are now experimentally available, including silicene[1,2] and germanene[3,10]. In contrast, little experimental efforts have been devoted on growing stanene and no previous report has confirmed the existence of such a material structure. Some early MBE experiments in 1990s might possibly obtained the stanene-like structure[11,12], but no structural characterization with atomic-scale resolution was provided. Recently three-dimensional (3D) TI states were observed in epitaxially grown α-Sn(001) thin films on InSb substrate[13,14], however 2D stanene films have not been realized. In this work, for the first time we successfully grew stanene films on $Bi_2Te_3$(111) substrates by MBE. With the help of first principle calculations, the atomic and electronic structures of stanene on $Bi_2Te_3$ were determined by STM and ARPES. Our results advance the first major step towards the study of stanene experimentally.

Stanene is a 2D allotrope of Sn in graphene-like structure[4]. It is composed of a biatomic-layer of α-Sn(111), in which two triangular sublattices stack together, forming a buckled honeycomb lattice (Fig. 1a). The buckling of stanene, defined as the height difference between the top and bottom Sn atoms, depends slightly on chemical environment and is typically around 0.1 nm. Stanene can be nicely grown on $Bi_2Te_3$(111) substrate. Fig. 1b shows the typical reflected high-energy electron diffraction (RHEED) pattern after ultrathin Sn film deposition. According to the distance between the adjacent diffraction lines in the RHEED patterns before and after stanene deposition, we know that the in-plane lattice constant of ultrathin Sn film matches the substrate within the experimental uncertainty. Fig. 1c shows the RHEED intensity as a function of deposition time. One oscillation was observed. In fact, stanene film forms when we stopped the deposition near the peak position of the RHEED intensity as marked by the blue arrow in Fig. 1c. The $Bi_2Te_3$(111) substrate shows large terrace and steps of single-quintuple-layer (QL) height (~1nm) (Fig. 1d). Fig. 1e shows the STM topography on the terrace when we stopped the deposition at the time marked by the black arrow in Fig. 1c. Under this coverage, several layers of Sn are observed. The measured height line profile in Fig. 1f (corresponding to the black line in Fig. 1e) shows that the height between the first, second and third layers of Sn is the same. It is about $0.35\pm0.02$ nm and is consistent with the space distance between biatomic layers along the <111> direction in $\alpha$-Sn. This indicates the Sn film growing as biatomic-layer structure. Large-scale STM topography of stanene film (the first biatomic layer) is presented in Fig. 1g. Compared with the topography of the substrate (Fig. 1d), the overall surface morphology is unperturbed in the large scale after stanene growth. Sharp $Bi_2Te_3$ steps with 1 nm height were still observed (Fig. 1g and 1j). Fig. 1h shows the zoom-in STM image of stanene film. The film has a similar surface morphology as other ultrathin epitaxial films. About 90% of substrate is covered by stanene film and about 10% ~ 15% Sn atoms stay on the top as small islands. Triangular lattice of stanene film is clearly revealed in the atomic resolved STM image (Fig. 1i). In bulk or thick $\alpha$-Sn(111) films, a 2x2 reconstruction of surface Sn atoms was observed in STM[12]. In our ultrathin film we did not observe any reconstruction. The surface of stanene film is not perfectly flat on nanometer scale. The height line profile (Fig. 1k) (from the red line in Fig. 1i) shows a random height modulation of about 0.06 nm, which may be caused by the in-plane compressive stress or hydrogen adsorption. Similar modulation was also observed in previous study on thick Sn(111) films[15] and SnTe(111) films[16]. The biatomic-layer structure of the stanene film is further confirmed in Fig. 1l and 1m. At the edge of the island on the first Sn layer, we occasionally observed another Sn atomic-layer beneath the topmost Sn atomic-layer. The height difference shown in Fig. 1m (corresponding to the green line in Fig. 1l) between them is about 0.12±0.02 nm that nicely matches the distance between Sn atomic-layers in the biatomic-layer structure of stanene. We marked the position of atoms on these two atomic-layers in Fig. 1l. Very clearly, these two atomic-layers are A-B stacking like.

The relative position between Sn and Te atoms was also determined (See supplementary information). Based on the STM results in Fig. 1 and Fig. S1, we propose the atomic structure model for 2D stanene on $Bi_2Te_3$(111) in Figure 2. The green and orange balls present the bottom and top Sn atoms respectively in the biatomic-layer structure, respectively. The gray balls present the surface Te atoms of the substrate. As shown in Fig. 2(b), Sn atoms in the bottom atomic-layer (Sn-B) locate on the top of interstitial position of the Te layers. The Sn atoms in the top atomic-layer (Sn-A) locate at the hollow position (Fig. 2b, 2c).

The electronic structures of stanene on $Bi_2Te_3$ were studied by ARPES. Stanene is very thin, so the experimental ARPES spectra have the contributions both from the stanene and the substrate. In order to distinguish the energy bands of stanene from the ones of the substrate, incident photons with different energies as well as different polarizations were used to change the ARPES matrix elements[17]. Fig. 3a shows the ARPES spectra of $Bi_2Te_3$ along the K-Γ-K direction. Except the "V"-shaped surface bands (Dirac cone, indicated by white arrow in Fig. 3a) that cross the Fermi level at $k_F$ ~ 0.1 Å$^{-1}$ near the Γ point, all other valence bands (VB) are below the Fermi level. We marked the observed VB of the $Bi_2Te_3$ substrate near the Γ point by orange dashed lines in Fig. 3a. After the deposition of stanene, ARPES spectra changes dramatically (Fig. 3b). In Fig. 3b, Dirac cone of $Bi_2Te_3$ disappears and two hole bands around the Γ point appear (marked by blue dashed lines in Fig. 3b). The dispersion relation of the outer hole band was determined, which gives the Fermi vector $k_F \sim 0.3$ Å$^{-1}$. The inner hole band is less intense (the dispersion relation of the inner hole band is further determined by polarization dependence experiments discussed below). We think that these two hole bands originate from stanene. In addition, spectra of electron band like very close to the Γ point centered at about 150 meV below the Fermi level appear (indicated by white arrow in Fig. 3b). On the other hand, though the ARPES spectra of the VB of $Bi_2Te_3$ become weak and blurred due to the coverage of the stanene, the band dispersion of those bands are still resolvable in Fig. 3b. We overlaid the extracted $Bi_2Te_3$'s VB (orange dashed lines) from Fig. 3a on Fig. 3b. In order to match the spectra in Fig. 3b, orange dashed lines are shifted to higher binding energy by about 200 meV. Obviously, lots of spectral features below the Fermi level in Fig. 3b can be nicely identified as contributions from the substrate except the two hole bands. The shift of the VB of the $Bi_2Te_3$ to higher binding energy after the stanene growth indicates an electron transfer from the stanene to the $Bi_2Te_3$ substrate. Furthermore, compared with previously published $Bi_2Te_3$'s ARPES spectra[18-20], we assign the electron-like bands very close to the Γ point coming from the $Bi_2Te_3$'s bulk conduction bands (CB).

Fig. 3c shows the constant-energy plot of stanene/$Bi_2Te_3$ in momentum space at the Fermi energy (Fermi surface mapping) over six two dimensional Brillouin zones (BZs). Consistence with the observed band dispersions in Fig. 3b, the measured Fermi surface of stanene/$Bi_2Te_3$ is very different from that of the $Bi_2Te_3$ substrate (The Fermi surface of the $Bi_2Te_3$ substrate is presented in the supplementary information). In the first BZ (labeled by "I" in Fig. 3c), the Fermi surface shows three-fold symmetry in intensity, which is related to the three-fold symmetry of the biatomic-layer structure of stanene. Furthermore, there are Fermi surfaces (regardless the intensity variation in different BZs) around each Γ point in six BZs, which agrees with hexagonal Bravais lattice of stanene as found in STM image in Fig. 1i. Figure 3d shows two ARPES spectra along two momentum directions (Γ-M and Γ-K) as labeled by the yellow dashed lines in Fig. 3c. No other bands cross the Fermi level except two hole bands. Both at M and K points, all energy bands are at least 500 meV below the Fermi level.

The dispersion of the inner hole band are further determined by the polarization dependent experiments. Fig. 3f and 3g present the ARPES spectra near the Γ point under *p* and *s* polarization geometries, respectively. Fig. 3e shows the sketch of *p* and *s* polarization geometries. Under *p*-polarization, we observed two hole bands (marked by white dotted lines) crossing the Fermi level. Additionally, $Bi_2Te_3$'s CB and VB have high intensities (indicated by white arrows). Those $Bi_2Te_3$'s bands are strongly suppressed under *s*-polarization (Fig. 3g), where two hole bands originated from stanene remain and the inner hole band is enhanced. Consequently, the dispersion of the inner hole band near the Fermi level was determined. Figure 3h presents the complete AREPS spectra of stanene/$Bi_2Te_3$(111) along the M-Γ-K-M direction. We marked the experimental low energy bands that were not observed on pure $Bi_2Te_3$(111) by blue dotted lines in Fig. 3h. We think those bands should originate from stanene.

One thing we need to point out is that the contributions of the $Bi_2Te_3$ substrate were determined experimentally. We didn't use the published LDA calculations of $Bi_2Te_3$[21] as the references to distinguish the contributions from stanene because the LDA bands are not completely consistent with experiments as revealed recently in ARPES experiments[22]. For example, one misleading band in Fig. 3h is the third hole band below the Fermi level near the Γ point (marked by green dased lines in Fig. 3h) that was also observed in Fig. 3a (high resolution AREPS spectra are also shown in supplemental information Figure S5). According to LDA calculations[21, 22], this hole band does not belong to $Bi_2Te_3$, which would cause the uncertainty to determine the bands of stanene. Recent high accuracy GW calculations[22], which reproduce more subtle details in the experimentally observed bands though the agreement between experiments and calculations is still not excellent, predict similar hole band for bulk $Bi_2Te_3$[22]. Independent of calculations, we also carried out in situ surface potassium deposition to dope electrons to the surface. We found that the Fermi vector of the two hole bands of stanene becomes smaller with the surface electron doping, while the third hole band barely changes, which means that electrons was doped into stanene's bands not into the third hole band (See supplementary information Fig. S2). This finding also indicates that the third hole band does come from the $Bi_2Te_3$ substrate.

To help understand the experimental results, we performed density functional theory (DFT) calculations for stanene on $Bi_2Te_3$(111). The free-standing stanene would have a Dirac cone at the K (and K') point if excluding the spin-orbit coupling (SOC) and have a band gap when including SOC[4]. Stanene on $Bi_2Te_3$(111) has a commensurate surface lattice. Herein the surface lattice constant of $Bi_2Te_3$(111) (4.383 Å) is smaller than that of the free-standing stanene (4.676 Å from DFT calculations)[4]. The substrate thus applies a compressive strain on stanene. For a free-standing stanene, the compressive strain increases the magnitude of buckling from 0.85 Å to 1.09 Å. Meanwhile, the valence band at the Γ point shifts upwards with respect to the conduction band at the K point, resulting in a 2D TI-to-metal transition. For stanene on $Bi_2Te_3$(111), we found that a single atomic Sn layer would bind preferably at the fcc site that is much more stable than at the top site, supporting the experimental model. Interestingly, different binding sites of stanene give very similar geometric and electronic properties, possibly due to the weak stanene-substrate interaction (the calculated binding energy is ~0.1 eV per surface unit cell). For comparison, we considered a structure in which the bottom (top) Sn atoms are located at the hollow (fcc) site of the substrate (Fig. 4a) as proposed by experiment. The optimized distance between the top atomic-layer of stanene and the Te layer of the substrate is 4.42 Å. Upon adsorption the buckling of stanene increases slightly to 1.14 Å, which agrees nicely with our experimental result (1.2 Å). In the band structure of the combined system (Fig. 4b), the contribution from stanene is clearly visualized. Stanene under compressive strain remains metallic when placing on $Bi_2Te_3$(111). The metallic states of stanene, however, can be gapped, for instance, by chemical functionalization[4,7]. Chemically saturating all the $p_z$ orbital of stanene could result in QSH states, induced by a band inversion at the Γ point[4]. Moreover, selectively saturating half of the $p_z$ orbital could introduce ferromagnetism in stanene, induce band inversion for one spin channel, and thus result in novel QAH states[7]. DFT-PBE calculations typically underestimate the binding strength for weak adsorption systems. To check the influence, we artificially decreased the adsorption height by 0.5 Å, and found a moderate energy increase (by 0.3 eV) and negligible changes in the band structure (except larger band splitting at the K point). All these results, together with the fact that no chemical bond is formed between the two subsystems, consistently suggest a weak-coupling picture for stanene on $Bi_2Te_3$(111).

Despite the weak coupling, our electronic structure calculations revealed some important features induced by interfacing stanene and $Bi_2Te_3$. First, the conduction bands of $Bi_2Te_3$(111) get partially occupied upon stanene growth, indicating an electron transfer from stanene to the substrate, in agreement with the ARPES data. As a result, the Fermi level pins two hole bands of stanene with bulk CB of $Bi_2Te_3$(111) at the Γ point, as observed by ARPES. Secondly, the existence of the substrate breaks the inversion symmetry of stanene, leading to band splitting. The splitting is small at the Γ point but noticeable at the K point (Fig. 4b). Thirdly, the band gap of stanene at the K point depends sensitively on environment, such as the substrate or adsorbates (like hydrogen). The associated electronic states at K point are mainly contributed by the $p_z$ orbital of Sn. This $p_z$ orbital is chemically active due to its unsaturated nature, and is thus easily affected by external environment. Specifically, the band gap of stanene at the K point gets enhanced upon adsorption and would increase if decreasing the adsorption distance. An extreme value of 6 eV is obtained in a hydrogenated stanene with all the $p_z$ orbital fully saturated. Since hydrogen ubiquitously exists in the growth process, the band structure of stanene at the K point is expected to vary significantly with growth conditions.

We compared the experimental determined bands with DFT calculations in Fig. 4c. Energy bands above the Fermi level (red dots in Fig. 4c) were obtained using *in situ* surface potassium deposition (surface electron doping) (See supplementary information). The calculated bands in Fig. 4c only include the contributions from stanene. In very good agreement with DFT calculations, at the Γ point, two hole bands cross the Fermi level and the Fermi vectors ($k_F$) nicely match the calculations. Even though along Γ--M direction, the experimental band dispersion agrees very well with DFT results, the bandwidth of the outer hole-like band is slightly smaller than calculations. Along Γ--K direction, there is obvious discrepancy between DFT calculations and experimental results. Near the K point, bands originated from the $p_z$ orbital were not observed near the Ferm level. In some sense, this discrepancy is expected. Due to the unsaturated nature of the $p_z$ orbital, adsorbates can easily bind on stanene, shifting the $p_z$ orbital of stanene away from the Fermi level. The disappearance of $p_z$ orbital might be attributed to the existence of adsorbates (like hydrogen) in the experiment that is not included in the calculation. However, this does not affect the band features at the Γ point. The easily influenced $p_z$ orbitals of stanene make the bands at the K point sensitively dependent on the environment. The material trait on the other hand offers great tunable functionalities and new possibilities for the stanene research. Many approaches have been proposed to create chemically distinct derivatives of stanene, for instance, by chemical functionalization[4], adding more Sn atoms[23], increasing the layer thickness[24], or forming backbone bonds with substrate[7]. These derivative materials support various novel quantum states including QSH and QAH.

In summary, we have successfully grown ultrathin Sn films with 2D stanene structure on the substrate of $Bi_2Te_3$. The atomic and electronic structures of epitaxial stanene films were experimentally determined, which agree well with the theoretical predictions. The hole pockets of stanene observed experimentally at the Γ point, with the help of the Rashba effect due to the breaking of the inversion symmetry on the substrate, could possibly support topological superconductivity[6]. Furthermore, the metallic states of stanene can be gapped, for instance, by chemical saturation of its $p_z$ orbital. With gapped bulk states, the stanene system can support novel QSH and QAH states as proposed in Refs. [4,7]. Our results open a door for studying properties of 2D stanene in the future.

**Methods**

Experiments:

Bi$_2$Te$_3$ thin films and bulk single crystals are used as substrates. Bi$_2$Te$_3$(111) films up to 40nm were grown by MBE method on the Si (111) wafer. Bulk single crystals were grown by the modified Bridgman method. Single crystals were cleaved in situ at 30 K, resulting in shiny, flat, and well-ordered surfaces. High purity Sn (99.999%) was evaporated from the effusion cells. The deposition rate is about 0.4 monolayer/min. The substrate was kept at room temperature. The thickness and growth mode of Sn films was monitored by RHEED and STM. The sample temperature was kept at 30 K during ARPES measurement. The STM images were acquired at 77K and 4K. ARPES measurements were performed with 60–90 eV photons at Advanced Light Source beamlines 4.0.3 using Scienta R4000 analyzer with base pressures better than $5 \times 10^{-11}$ Torr. Energy resolution is better than 15 meV and angular resolution is better than 1% of the Brillouin zone. Surface electron doping was performed by *in situ* potassium deposition using Alkali metal dispenser (SAES Getters).

Calculations:

First-principle calculations based on density functional theory (DFT) were performed using the projector-augmented-wave potential and the exchange-correlation functional of the Perdew-Burke-Ernzerhof (PBE) generalized gradient approximation, as implemented in the Vienna *ab initio* simulation package. The Bi$_2$Te$_3$ surface was modeled by a periodic slab, including four QLs with the bottom three fixed at the bulk crystal structure, using the experimental lattice constants of $a$ = 4.383 Å and $c$ = 30.487 Å. The slab calculations used a plane wave basis with an energy cutoff of 200 eV (250 eV for systems having hydrogen), a 11×11×1 Monkhorst-pack $k$ grid, a vacuum layer over 10 Å, and dipole corrections between periodic images. Force convergence criteria of 0.01 eV/Å was selected for structural optimization, and the spin orbit coupling was included in electronic structure calculations.

**Acknowledgements**

The work in SJTU was supported by the Ministry of Science and Technology (MOST) of China (2013CB921902, 2011CB922200,), NSFC (11227404, 11274228). The work in Stanford is supported in part by the Department of Energy, Office of Basic Energy Sciences, Division of Materials Sciences and Engineering, under contract DE-AC02-76SF00515 and by FAME, one of six centers of STARnet, a Semiconductor Research Corporation program sponsored by MARCO and DARPA. D.Q. acknowledges additional support from the Top-notch Young Talents Program and the Program for Professor of Special Appointment (Eastern Scholar) at Shanghai Institutions of Higher Learning. The Advanced Light Source is supported by the Director, Office of Science, Office of Basic Energy Sciences, of the US Department of Energy under Contract DE-AC02-05CH11231. The work is also supported by ENN.

**Author contributions**

F.F.Z. and C.W.J. conducted the experiments with the help of D.Q. and C.L.G.. Y.X. conducted the calculations. D.Q., Y.X., S.C.Z. and J-F.J. designed the experiments and provided financial and other supports for the experiments. D.Q., C.L, Y.X., D.D.G., C.H.L. and J-F.J. analysed the data. D.Q., Y.X., and J-F.J. wrote the paper.

**Competing financial interests**
The authors declare no competing financial interests.

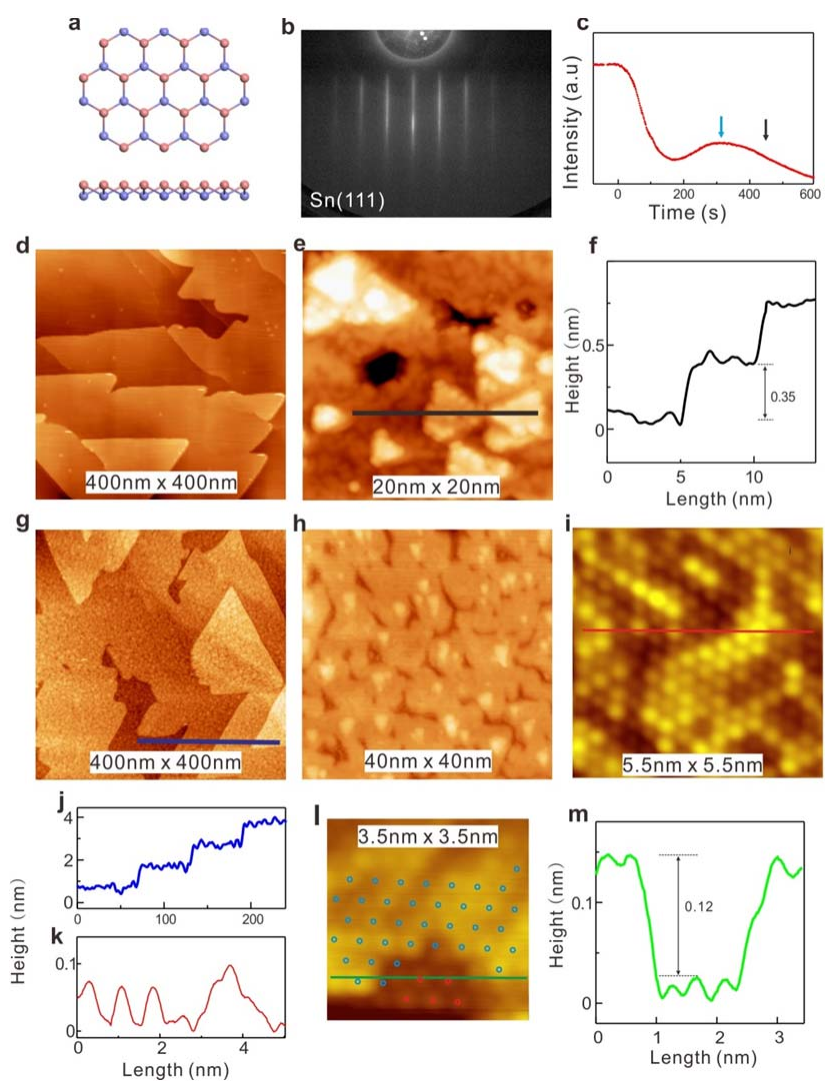

**Figure 1| Atomic structures of stanene on Bi$_2$Te$_3$. a**, Crystal structure of stanene from the top (side) view [upper (lower)]. **b**, RHEED pattern of stanene film. **c**, RHEED intensity as a function of growth time. The blue arrow marks the deposition time for stanene. STM topography of **d**, Bi$_2$Te$_3$(111) and **e**, Sn films of more than single biatomic layer coverage. The corresponding deposition time is marked by the black arrow in **c**. **f**, Height line profile in **e**. **g**, Large-scale STM topography of stanene film. **h**, Zoom-in STM image of stanene. **i**, Atomically resolved STM image of stanene. **j** and **k,** Height line profiles in **g** and **i**. **l**,

Atomically resolved STM image of top and bottom atomic layers of stanene. Blue dots mark the lattice of the top Sn atoms. Red dots mark the lattice of the bottom Sn atoms. Two lattices do not coincide. **m**, Height line profile in **l**.

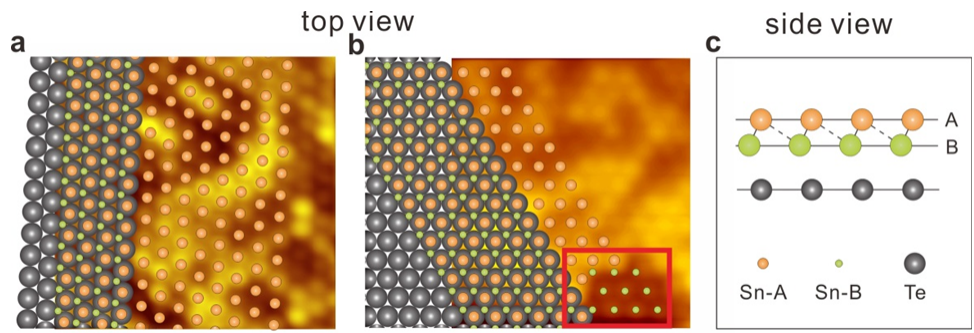

**Figure 2| Atomic structure model for the 2D stanene on Bi$_2$Te$_3$(111). a**, Top view of the top Sn atoms. **b**, Top view of both the top and bottom Sn atoms. (c) Side view.

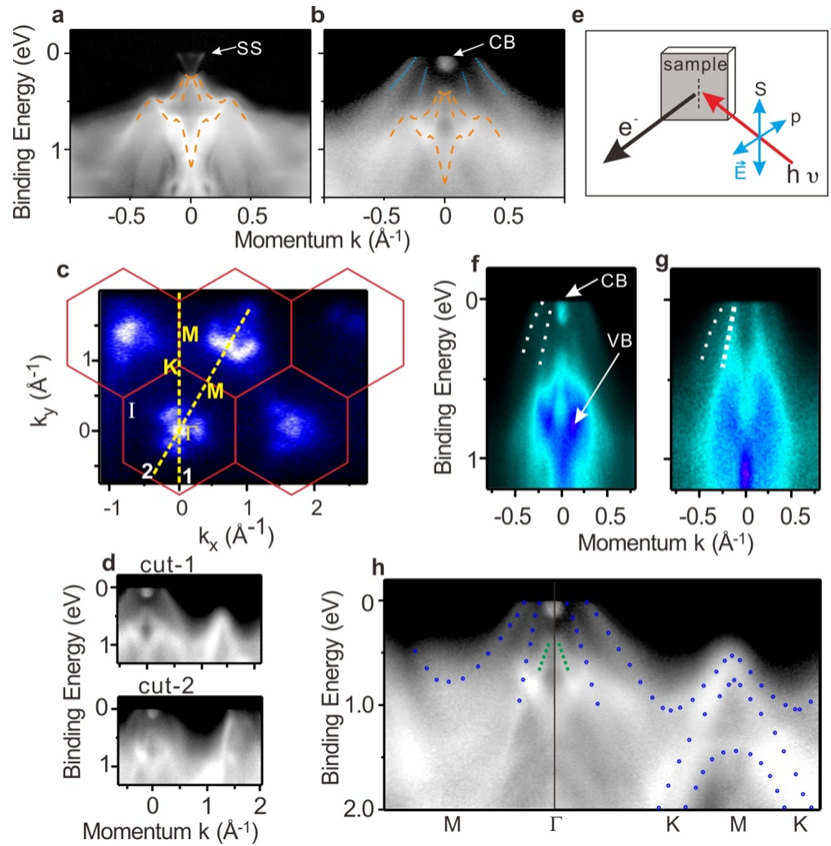

**Figure 3| Electronic structures of stanene film.** ARPES spectra of **a**, Bi$_2$Te$_3$(111) and **b**, Stanene on Bi$_2$Te$_3$ along K-Γ-K direction. The orange dots mark the band dispersions of Bi$_2$Te$_3$. The blue dashed lines mark the hole bands of stanene. **c**, Fermi surface mapping in large momentum space. **d**, ARPES spectra along two momentum directions as marked by yellow lines in **c**. **e**, Sketch of two light polarization used in experiments. ARPES spectra taken under **f**, *p*-polarization and **g**, *s*-polarization. White dotted lines mark the hole bands of stanene. **h**, ARPES spectra along Γ-M-Γ-K-M-K directions. Blue dotted lines mark the experimental electronic bands of stanene. Green dashed lines mark one of the hole bands of Bi$_2$Te$_3$(111).

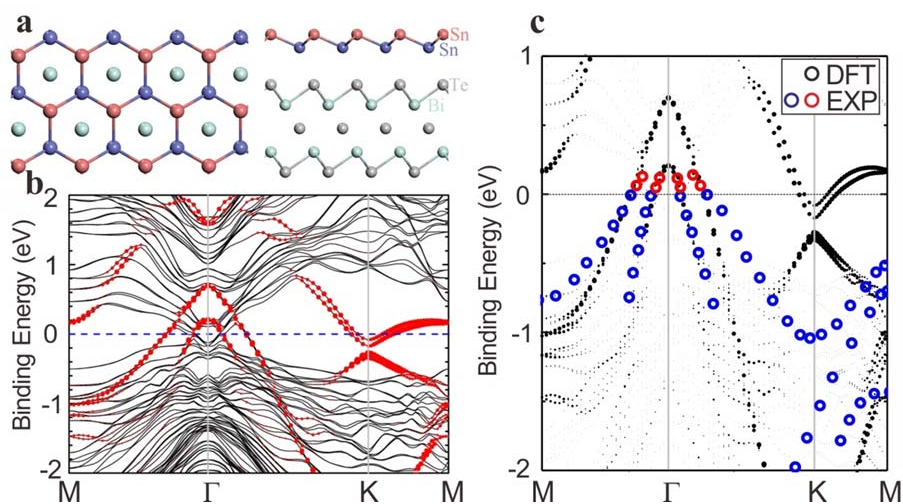

**Figure 4| Comparison between DFT calculations and experiments**. **a**, Crystal structure from the top (side) view [left (right)] and **b**, band structure of stanene on $Bi_2Te_3$(111). Only the top quintuple layer of the $Bi_2Te_3$(111) surface is shown here. The red dots on the bands illustrate the contribution from stanene. **c**, Comparison of experimental bands with DFT calculation of stanene/$Bi_2Te_3$. Red dots above the Fermi level are obtained by *in situ* potassium deposition that provides the film with electrons.